# HOW AN AUTOMATED GESTURE IMITATION GAME CAN IMPROVE SOCIAL INTERACTIONS WITH TEENAGERS WITH ASD

L. N. VALLÉE*, S. M. NGUYEN, C. LOHR, I. KANELLOS, O. ASSEU

*Abstract*— With the outlook of improving communication and social abilities of people with ASD, we propose to extend the paradigm of robot-based imitation games to ASD teenagers. In this paper, we present an interaction scenario adapted to ASD teenagers, propose a computational architecture using the latest machine learning algorithm Openpose for human pose detection, and present the results of our basic testing of the scenario with human caregivers. These results are preliminary due to the number of session (1) and participants (4). They include a technical assessment of the performance of Openpose, as well as a preliminary user study to confirm our game scenario could elicit the expected response from subjects.

## I. INTRODUCTION

Autism spectrum disorder (ASD) is a neurodevelopmental condition that includes issues with communication and social interactions. Information and communication technologies have contributed to the social and cognitive stimulation of children with ASD. This seems to be partly because these are more comfortable with predictive and repetitive behaviors [1], which can be implemented through algorithms. In particular, artificial intelligence algorithms are used in various fields, from language or gesture recognition to image classification [2]. These functions allow for a system to interact with a human being.

Furthermore, robots seem to prove useful with autistic children because children with ASD show particular interest in interacting with robots [3]. It is a promising field to use robots to perform gesture imitation learning [4] since the imitation process is a pillar for learning and interacting. Works mentioned in [5] demonstrate that autistic children are able to imitate.

Through imitation practice it appears possible to improve imitation abilities and even reduce the degree of autism, as shown in [6] where the experimentation consisted of human caregivers performing several simple imitation games with the children. The described effects were observed when the child was asked to imitate a caregiver or during "mirroring" interactions when he was imitated by a caregiver. In order to evaluate the quality of the imitation, the scale defined by Nadel [5] was used.

That work is interesting, but does not build on more modern tools, such as imitation learning algorithms or robots.

In [7, 8], it is shown how an interactive robot could bring positive results in terms of social interactions with autistic children. Several explanations such as predictability, simplicity in the movements, absence of implicit communication messages, or simpler face expressions than those of human beings (see Figure 1), have been proposed.

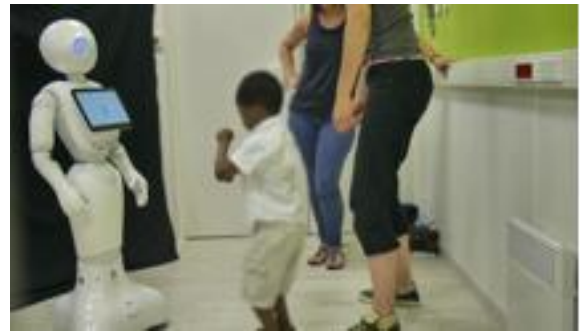

**Figure 1:** Interaction between a child with ASD and Pepper robot [3]

In this paper, we propose an automated gesture imitation game aiming at improving social interactions with autistic teenagers and preteens. As steps on the path to this purpose, skeleton detection is tested on teenagers and preteens with ASD, through the different phases of the game.

The work presented here is a pilot showing observations from particular cases.

The following section of this paper describes the game structure and methodology before simulation results are presented and discussed in section III.

Section IV contains a short summary of the work as well as some perspectives.

*The present research has not benefited from any financial support.

L. N. Vallée is with the ESATIC (École Supérieure Africaine des Technologies de l'Information et de la Communication), LASTIC, Abidjan, Côte d'Ivoire (e-mail : linda.vallee@esatic.edu.ci)

S. M. Nguyen is with the Institut Mines Télécom Atlantique, Brest, France (e-mail: nguyensmai@gmail.com) and with the Flowers Team of U2IS ENSTA Paris, Institut Polytechnique de Paris & INRIA.

C. Lohr is with the Institut Mines Télécom Atlantique, Brest, France (e-mail: christophe.lohr@imt-atlantique.fr)

I. Kanellos is with the Institut Mines Télécom Atlantique, Brest, France (e-mail: ioannis.kanellos@imt-atlantique.fr)

O. Asseu is with the ESATIC (École Supérieure Africaine des Technologies de l'Information et de la Communication), LASTIC, Abidjan, Côte d'Ivoire (e-mail : olivier.asseu@esatic.edu.ci)

## II. METHODOLOGY

### A. Test environment and global game structure

Work sessions are regularly held with autism professionals, autistic teenagers and their parents, with the aim of keeping connected with real-life challenges and needs. This is crucial in order to set up an adequate environment and propose a game that will actually be usable.

The test environment must be quiet and free of distractions.

In order to ease interactions with the participants, autism professionals recommended that our gesture imitation game would start with greetings and pairing phases, followed by several imitation modules: for instance, one induced, one spontaneous, with or without objects.

### B. Skeleton detection

A skeleton detection method is needed from the beginning and all through the game. The Tensorflow implementation of the Openpose algorithm [10] was tested. Tensorflow is a library used to train and execute neural networks for element classification like in gesture recognition.

The Openpose algorithm allows for the detection, through the computer camera, of the skeleton of the participant frame by frame. It was chosen over the use of a Kinect because it is more robust to occlusions and yaw rotations.

Figure 2 displays a schematic representation of the human body in Openpose.

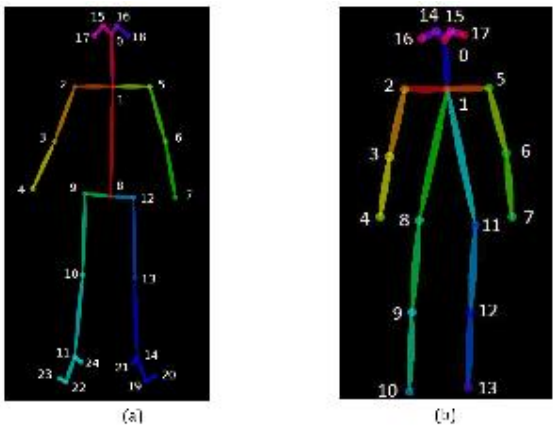

**Figure 2:** Schematic human body representation [9]

### C. Characteristics of the participants

Sessions were held with teenagers and preteens with ASD. Consent documents were previously signed by the responsible person.

- Assistant A was seated on a chair near the computer table running the pose detection tool.
- Assistant B was in charge of video recording the whole scene from a corner of the room.
- Assistant C (who shall later be replaced with the interactive robot) was standing in the room, waiting to attend the teenager with ASD.
- Assistant D was in charge of video recording the scene from Assistant C's point of view.
- Assistant E was outside the room with the teenager.

Characteristics (biological and neurodevelopmental ages, CARS scores) of the four teenagers and preteens are displayed below:

TABLE 1. CHARACTERISTICS OF THE PARTICIPANTS

| Identifier of participant | Biological age | ND (neuro developmental age | CARS* score |
|---|---|---|---|
| F | 18 | 9 | 33 |
| G | 14 | 4 | 46 |
| H | 13 | 5 | 38 |
| I | 12 | 1/2 | 47 |

*Childhood Autism Rating Scale – used to determine the degree of autism.

### D. Detailed game scenario

Following discussions with child psychiatrists and educators, the following phases and parameters were identified.

Phases:
- Greetings
- Pairing
- Imitation of sport movements
- Closing

Parameters observed:
- Head/body orientation
- Smile
- Joint attention
- Imitation attempts

Details about phases and evaluation methods are presented thereafter. Please see Appendix for a flowchart.

**a) Greetings phase**
The robot / Assistant C welcomes the participant with a smile and introduces herself. In case the participant is verbal, the robot / Assistant C asks him or her for their name. Finally, the robot / Assistant C offers her hand to greet then waits for 30 seconds.

- Success (3): the participant reaches out to the robot / Assistant C to greet.
- Intermediate (2): the participant does not reach out to the robot / Assistant C to greet but displays signs of interest. Are considered as such: head orientation towards the robot / Assistant C and/or smile.

- Failure (1): the participant does not show any interest towards the robot / Assistant C.

Note: we do not differentiate here between participants reacting immediately or later during the 30 second interval.

**b) Pairing phase**
The robot / Assistant C smiles (or displays lighting/sounds suggesting happiness), tells the participant how happy she is to be with him or her, leads a gentle talk, then invites (holding hand out) him or her to do some sport together and waits 30 seconds for a reaction.

- Success (3): the participant holds the robot / Assistant C's hand
- Intermediate (2): the participant does hold the robot / Assistant C's hand but displays signs of interest. Are considered as such: head orientation towards the robot / Assistant C and/or smile.
- Failure (1): the participant does not show any interest towards the robot / Assistant C.

Note: we do not differentiate here between participants reacting immediately or later during the 30 second interval.

**c) Imitation phase**
The following physical movements were successively proposed:
- Raising arms towards the sky to stretch the body
- Extending arms on each side of the body then bending forward
- Extending arms forward then bending towards the toes

Those movements were initially proposed without holding objects, then if needed, with colored balls. Spontaneous imitation was hoped for, then imitation was provoked.

- Success (3a): the participant tries to imitate the robot / Assistant C and the gesture recognition algorithm detects that the imitation is successful.
- Intermediate (2a): the participant tries to imitate the robot / Assistant C, but the gesture recognition algorithm detects that the imitation is unsuccessful.
- Failure (1a): the participant does not seem to try imitating the robot / Assistant C.

In case of failure, the robot / Assistant C would start imitating the participant in order to create a social interaction based on the recognition of being imitated.

- Success (3b): the participant would react positively to being imitated (smiling, laughing, diversifying movements or changing rhythm in order to encourage imitation)
- Intermediate (2b): the participant does not express explicit positive reactions to the imitation, but nevertheless shows increased attention towards the robot / Assistant C.

- Failure (1b): the participant does not show any interest towards the robot / Assistant C.

In the final game setup with the robot, the objective of this phase will be that there is an actual imitation (detected by the algorithm) and a social relationship is established between the teenager with ASD and the robot, here represented by Assistant C.

During the imitation phase, there are some necessary differences of positioning between both setups.
In the current setup with Assistant C and Openpose running on a computer with an internal camera, Assistant C positions herself on the participant's side and both are in front of the camera. But in the final setup with the robot, the participant will stand in front of the robot and both will interact based on the implemented algorithm.

**d) Closing phase**
Thanking the participant and accompanying him or her towards Assistant E.

## III. USER TEST RESULTS AND DISCUSSION

With this methodology, we carried out a pilot study with one session and four participants. The goal was to assess the validity of our scenario and the pose detection tool. No clinical assessment was performed.

Below are some Openpose captures of the sessions with the participants, along with descriptions and analysis for the body parts not detected.

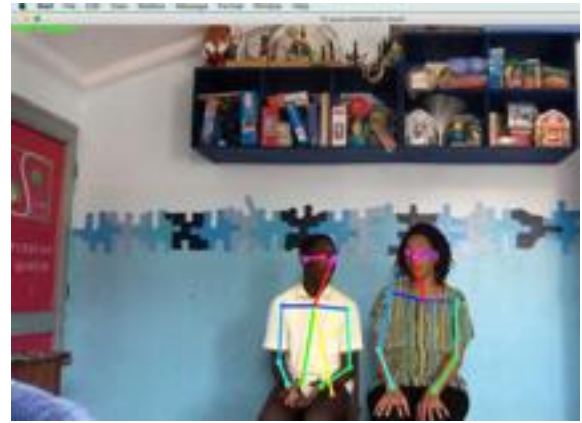

**Figure 3:** Capture of the pairing phase with Participant F

On Figure 3, Participant F and Assistant C are seated and talking during the pairing phase.
Due to the short distance between the camera and the participants, only the upper part of the body is visible.
Participant F's body segments are correctly detected.
Assistant C's shirt seems to impede proper detection of the torso. Further experimentation would be needed in order to determine the root cause. In the following captures, we will not highlight this aspect.

On Figure 4 below, which was captured during the imitation phase, Participant and Assistant C are standing in front of the camera. Openpose partially detected their skeletons. Due to the short distance between the camera and the participants, lowest body segments are not visible. The left leg of Participant F was not detected, maybe because the hand occluded part of it and the knee was not visible.

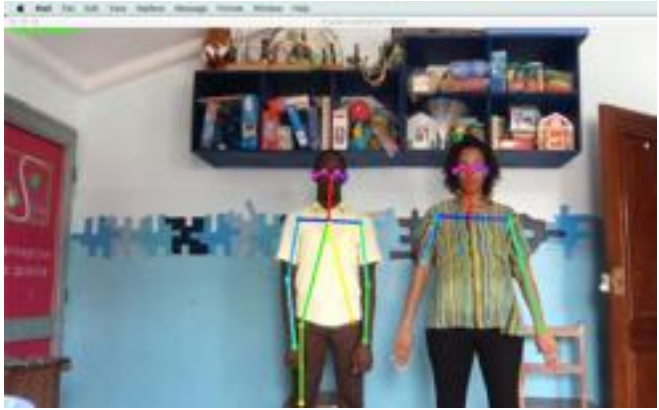

**Figure 4:** Capture of the imitation phase with Participant F

On Figure 5, Participant G and Assistant C are standing in front of the camera. Assistant C raises the left arm and Participant G the right one. Free hands are being held. The imitation of the first movement (both arms raised towards the sky) is therefore partial, mainly because Participant G needed to hold Assistant C's hand.

Furthermore, skeleton detection does not properly work for several reasons, among which:
- Participant G is wearing a skirt, which impedes proper detection of the legs;
- Light saturation on the top-right corner, causing the non-detection of the raised arms;
- Body occlusion where hands are joined.

The non-detection of the left leg of Assistant C may be due to the strong lighting above.

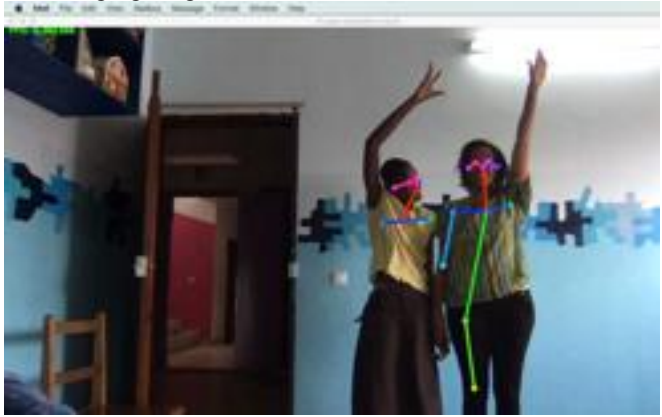

**Figure 5:** Openpose capture from the imitation phase with Participant G (1)

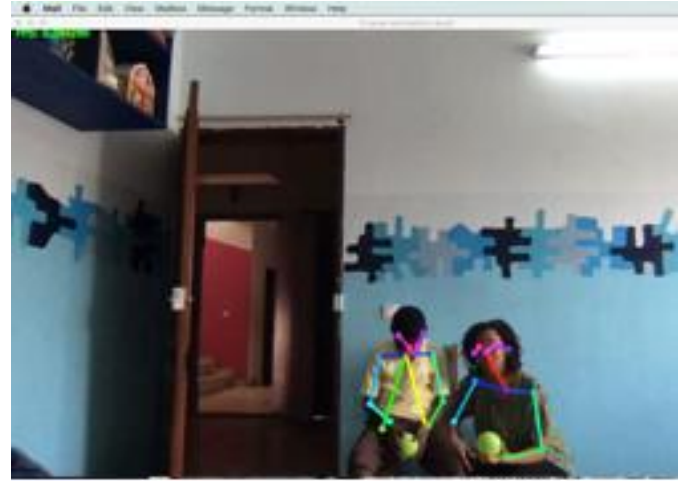

**Figure 6:** Openpose capture from the pairing phase with Participant H

The pairing phase with Participant H was intricate.
He had just woken up and would not enter the room. Assistant C had to get out of the room and spend some time with him in the hall. He would not greet, neither look at her nor leave the wall. After about one minute of gentle talk, Participant H accepted to follow Assistant C into the room but immediately wanted to sit down, head bended down, as shown on Figure 6. He would not react to invitations to stand up and do some sport together. It was then decided to use the colored balls to get his attention.
This session required much patience and adaptability. Participant H finally imitated some of Assistant C's movements, as shown in Figure 7. It will be interesting to see how Participant H will react to an interactive robot implementing a gesture imitation game.

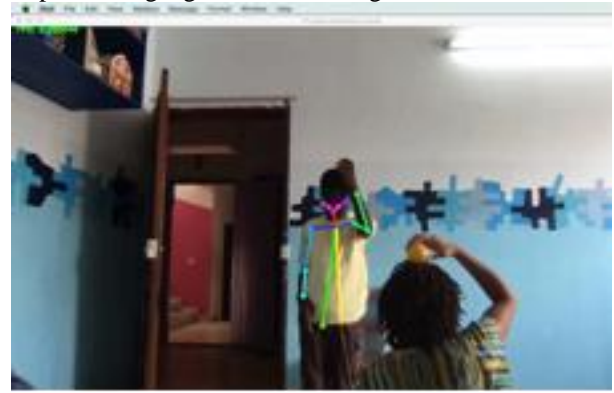

**Figure 7:** Openpose capture from the imitation phase with Participant H

Figure 8 below displays a capture of the pairing phase with Participant I, where Assistant C is trying to attract the participant's attention by running on the same spot.
Assistant C's body is almost not detected by Openpose, which could be explained by the fact that she is from side-on and in movement.
Participant I's skeleton is fully detected, in spite of the fussy background.

Surprisingly it is noted that a miniature "skeleton" is detected on the motifs of the wall. This type of false-positive could be filtered out through a heuristic approach.

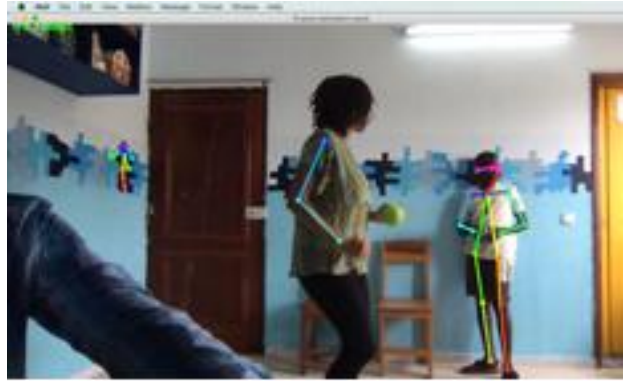

**Figure 8:** Openpose capture from the pairing phase with Participant I

Table 2 summarizes the results for each participant, per phase (greetings, pairing, imitation) and globally. Coded results are based on the scale suggested in section II. In addition, relevant comments are included about the type of imitation achieved (with or without objects) and the results in terms of social interactions (established or not, immediately or progressively).

TABLE 2. RESULTS OF THE RESPONSES OF PARTICIPANTS FOR EACH PHASE OF THE INTERACTION

| Participant ID | greetings | pairing | imitation | Comments |
|---|---|---|---|---|
| F | 3 | 3 | 3a | CARS 33, imitation without objects, good social interaction rapidly established |
| G | 3 | 3 | 2a | CARS 38, imitation without objects, good social interaction progressively established (initial shyness, smiles and hugs at the end) |
| H | 2 | 3 | 2 | CARS 47, imitation with objects, H had just woken up, medium social interaction progressively established, took more than one minute to get H to try imitating. |
| I | 3 | 3 | 3b | CARS 47, recognition of being imitated (singing), good social interaction progressively established (giggles), J was distracted because of lunchtime. Very mobile. |

Also, a simple video was created to display relevant extracts from greetings, pairing and imitation phases: https://youtu.be/Ueq4gULVd_c

Within the framework of our gesture imitation game for teenager and preteens with ASD, skeleton detection can be performed using the Openpose algorithm.

The environment settings, the participants' clothing as well as the body postures included within the game, of course must be carefully chosen in order to make sure that those are correctly detected.

For instance, lighting should be sufficient and uniform, background not too fussy and body occlusion should be limited. Also, postures from front or three-quarters are preferred. Moreover, loose dresses, skirts and other wide clothes should be avoided.

## IV. CONCLUSION

In this work, after reviewing previous studies, several aspects of our methodology were presented: detailed structure of the gesture imitation game, technical method for body detection, several aspects of game implementation. Skeleton detection results from several game scenes were then analyzed. Subsequently, recommendations were made in order to adapt our gesture imitation game to several constraints identified with Openpose.

The complete serious game is to be developed and implemented through an interactive robot with the aim of improving social interactions with autistic teenagers.

The choice of the robot will include different considerations, among which size, degrees of freedom, presence of a screen.

## APPENDIX

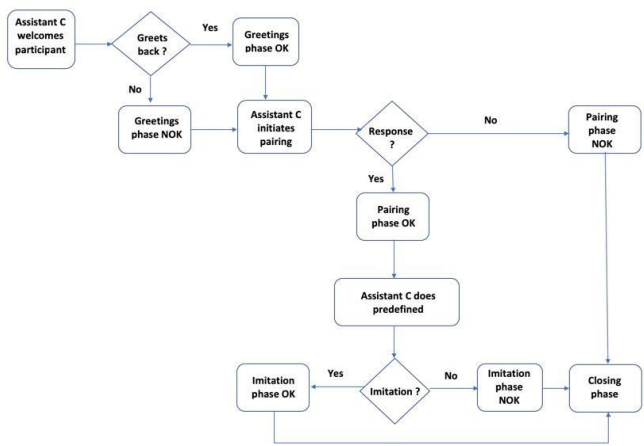

Flowchart for Imitation game scenario

ACKNOWLEDGMENT

Deep thanks to the CAMPSE, a center in Abidjan, Cote d'Ivoire, where young people with special needs are being accompanied by an experimented and dedicated team of educators and psychologists. Special thanks to the Director, Mrs Kieffoloh-Touré Miyala, Teacher Cynthia, Mr. Issa, Mr. Konan, Mr. Kouassi, and the wonderful children, preteens and teenagers who rewarded us with their genuineness.